\documentstyle[12pt]{article}
\let\Large=\large
\let\large=\normalsize
\begin{document}

\begin{flushright}
CAMS/99-02\\
hep-th/9903143
\end{flushright}
\vspace{.5cm}

\begin{center}
\baselineskip=16pt 
\centerline{{\Large {\bf
Anti-De Sitter BPS Black Holes}}} 
\centerline{{\Large {\bf
 in $N=2$ Gauged Supergravity}}} \vskip 1 cm {W. A. Sabra } \footnote{%
e-mail: {ws00@aub.edu.lb}} \vskip 1cm \centerline{\em Center for Advanced
Mathematical Sciences (CAMS), } \centerline{\em and} \centerline{\em Physics
Department, American University of Beirut, Lebanon}
\end{center}

\vskip 1 cm \centerline{\bf ABSTRACT} \vspace{1 cm} Electrically charged
solutions breaking half of the supersymmetry in Anti-De Sitter four
dimensional $N=2$ supergravity coupled to vector supermultiplets are
constructed. These static black holes live in an asymptotic $AdS_4$ space
time. The Killing spinor, $i. e.,$ the spinor for supersymmetry variation is
explicitly constructed for these solutions. \bigskip  \newpage

\section{Introduction}

Supersymmetric solutions of $N=2$ ungauged supergravity theory in four and
five dimensions have received lots of attention in recent years \cite{first,
black24, black25}. Such solutions play a fundamental role in probing the
(non)-perturbative phase transitions taking place among $N=2$ string vacua
and M-theory \cite{Strominger}. These solutions admit perturbative and
non-perturbative corrections and exhibit a rich structure due to less
supersymmetry and the special and very special geometries underlying $D=4$ and 
$D=5$ $N=2$ supergravity theories. BPS solutions provide a link between the
structure of the internal space, e.g.\ a Calabi-Yau threefold, and the
physical properties of the four-dimensional space-time. Singularities in
space-time could be related to special points in the internal space. Using
non-trivial space-time dependent solutions of $N=2$ supergravity, some
information can be obtained about the dynamical nature of the
non-perturbative topological transitions \cite{topological}.

Recently, there has been renewed interest in gauged supergravity theories in
various dimensions. It is motivated by the fact that the ground state of
these theories is anti-de Sitter (AdS) space-time and thus they may have
implications for the recently proposed AdS/conformal field theory (CFT) correspondence \cite{ibnsina}. This implies an equivalence of Type IIB string theory (or M-theory) on
anti-de Sitter (AdS) space-time and the conformal field theory (CFT) on the
boundary of this space. It is of special interest to address cases with
lower or no supersymmetry, in order to shed light on the nature of the
correspondence there. Supergravity vacua with less supersymmetry may have an
interpretation on the CFT side as an expansion of the theory around non-zero
vacuum expectation value of certain operators. Solutions with no
supersymmetry could also be viewed as excitations above the ground state of
the theory. This makes the study of black holes in gauged supergravity a
subject of current research interest. Moreover, it has been realised
recently that AdS spaces admit the so-called topological black holes which
have some unusual geometric and physical features \cite{top}.

The purpose of this paper is to describe static electric BPS black hole
solutions of D=4 gauged supergravity theories with vector supermultiplets.
These gauged supergravity theories admit the ${\hbox{AdS}}_{4}$ space-time
as a ground state. The analysis is general as we formulate our solutions in
terms of the holomorphic sections of $D$=4, $N$=2 supergravity. $N=4,8$ black holes appear as a special subclass for special choices of the prepotential of the $N=2$ theory. For the pure
supergravity case (i. e, no vector multiplets), Reissner-Nordstr\"om solutions have been discussed in \cite{romans}. Supersymmetric solutions have been obtained recently for the theory of $N=2$ supergravity in five
dimensions \cite{bcs} as well as for the $N=8$, $D=4$ theory \cite{jim}.
Also BPS topological black hole solutions for the pure $N=2$ supergravity
case have been constructed in \cite{klemm}. The spherically symmetric BPS
electric solutions can be obtained by solving for the vanishing of the
gravitino and gaugino supersymmetry variations for a particular choice for
the supersymmetry parameter. We will present here only static
non-rotating spherically symmetric electric solutions which break half of
supersymmetry and leave the rotating, magnetic and more general solutions
for a separate publication. The construction of our solutions relies very
much on special geometry of the $N=2$ supergravity theories. For this
reason, we review this subject and collect some formulae and expressions of $
N=2$ supergravity which will be important for the following discussion. Also
the general BPS solutions of $N=2$ black holes \cite{black24} in the theory
of ungauged $N=2$ supergravity with vector supermultiplets are briefly
discussed. Our conventions are collected in the Appendix. 

\section{Special Geometry and Black Holes in $N=2$ Supergravity}

Special geometry comes about when one couples vector supermultiplets to $N=2$
supergravity in four space-time dimensions (for a recent discussion and
references therein see \cite{v}) . The complex scalars $z^{A}$ of the vector
supermultiplets of the $N=2$ supergravity theory are coordinates which
parametrise a special K\"{a}hler manifold. Roughly, this is a
K\"{a}hler-Hodge manifold with a constraint on the curvature, $R_{A\bar{B}C%
\bar{D}}=g_{A\bar{B}}\,g_{C\bar{D}}+g_{A\bar{D}}\,g_{C\bar{B} }-C_{ACE}\,C_{%
\bar{B}\bar{D}\bar{L}}\,g^{E\bar{L}},$ here $g_{A\bar{B}}=\partial
_{A}\partial _{\bar{B}}K,$ is the K\"{a}hler metric with $K$ the K\"{a}hler
potential and $C_{ABC}$ is a completely symmetric covariantly holomorphic
tensor. K\"{a}hler-Hodge manifolds are characterised by a $U(1)$ bundle
whose first Chern class is equal to the K\"{a}hler class, thus, locally, the 
$U(1)$ connection can be represented by 
\begin{equation}
Q=-{\frac{i}{2}}\left( \partial _{A}Kdz^{A}-\partial _{\bar{A}}Kd\bar{z}%
^{A}\right).
\end{equation}
A definition of special K\"{a}hler manifold is given in terms of a flat $%
(2n+2)$ dimensional symplectic bundle over the K\"{a}hler-Hodge manifold,
with the covariantly holomorphic sections 
\begin{equation}
V=\left( 
\begin{array}{c}
L^{I} \\ 
M_{I}
\end{array}
\right) ,\qquad I=0,\cdots ,n\qquad D_{\bar{A}}V=(\partial _{\bar{A}}- {%
\frac{1}{2}}\partial _{\bar{A}}K)V=0,
\end{equation}
obeying the symplectic constraint 
\begin{equation}
i\langle V|\bar{V}\rangle =i(\bar{L}^{I}M_{I}-L^{I}\bar{M}_{I})=1.
\end{equation}
One also defines
\begin{equation}
U_{A}=D_{A}V=(\partial _{A}+{\frac{1}{2}}\partial _{A}K)V=\pmatrix{f_A^I\cr
h_{AI}}.
\end{equation}
where $D_{A}$ and $D_{\bar{A}}$ are the covariant derivatives\footnote{%
for a generic field $\phi ^{A}$ which transforms under the K\"{a}hler
transformation, $K\rightarrow K+f+\bar{f}$, by the $U(1)$ transformation $%
\phi ^{A}\rightarrow e^{-({\frac{p}{2}}f+{\frac{{\bar{p}}}{2}}{\bar{f}}
)}\phi ^{A},$ we have $D_{A}\phi ^{B}=\partial _{A}\phi ^{B}+\Gamma _{{AC}%
}^{B}\phi ^{C}+{\frac{p}{2 }}{\partial _{A}K}\phi ^{B}$, $D_{\bar{A}}$ is
defined in the same way}. In general, 
\begin{equation}
M_{I}={\cal {N}}_{IJ}L^{J}\ , \qquad\qquad h_{AI}={\bar{{\cal {N}}}}%
_{IJ}f_{A}^{J}\ .
\end{equation}
Special geometry can be defined in terms of the following differential
constraints 
\begin{eqnarray}
D_{A}V &=&U_{A},  \nonumber \\
D_{A}U_{B} &=&iC_{ABC}g^{C\bar{L}}{\bar{U}}_{\bar{L}},  \nonumber \\
D_{A}{\bar{U}}_{\bar{B}} &=&g_{A\bar{B}}{\bar{V}},  \nonumber \\
D_{A}{\bar{V}} &=&0
\end{eqnarray}
and \cite{v} 
\begin{equation}
\langle V,U_{A}\rangle =0.
\end{equation}
The K\"{a}hler potential can be constructed in a symplectic invariant manner
by defining the holomorphic sections $\Omega=e^{-\frac{K}{2}}V$ by 
\begin{equation}
K =-\log \left( i\langle \Omega |\bar{\Omega}\rangle \right) =-\log i(\bar{X}%
^{I}F_{I}-X^{I}\bar{F}_{I}).
\end{equation}
Moreover, special geometry implies the following relations 
\begin{eqnarray}
g_{A\bar{B}}&=&-i\langle U_{A}|\bar{U}_{\bar{B}}\rangle =-2f_{A}^{I}{%
\hbox{Im}} {\cal {N}}_{IJ}\bar{f}_{\bar{B}}^{J},  \nonumber \\
g^{A\bar{B}}f_{A}^{I}{\bar{f}}_{\bar{B}}^{J}&=&-{\frac{1}{2}}(\hbox{Im} 
{\cal {N}})^{IJ}-{\bar{L}}^{I}{L}^{J},  \nonumber \\
F_{I}\partial _{\mu }X^{I}-X^{I}\partial _{\mu }F_{I}&=&0.  \label{useful}
\end{eqnarray}

The $N=2$ supergravity action includes in addition to the gravitational
supermultiplets, a number of vector and hypermultiplets. Throughout this
work, the hypermultiplets are assumed to be constants. In this case, the
bosonic $N=2$ action is given by 
\begin{equation}
S_{N=2}=\int \sqrt{-g}\,d^{4}x\left( -{\frac{1}{2}}R+g_{A\bar{B}}\partial
^{\mu }z^{A}\partial _{\mu }\bar{z}^{B}+i\left( \bar{{\cal {N}}}_{IJ}{F}%
_{\mu \nu }^{-I}{F}^{-J{\mu \nu }}\,-\,{\cal N}_{IJ}{\it F}_{\mu \nu }^{+I}{%
\ }{\it F}^{+J{\mu \nu }}\right) \right)
\end{equation}
where $F^{\pm I\,\mu \nu }={\frac{1}{2}}\left( F^{I\,\mu \nu }\pm {\frac{i}{2%
}} \varepsilon ^{\mu \nu \rho \sigma }F_{\rho \sigma }^{I}\right)$.

The supersymmetry transformations for the chiral gravitino $\psi_{\alpha\mu}$
and gauginos $\lambda^{A \alpha}$ in a bosonic background of $N=2$
supergravity are given by 
\begin{eqnarray}
\delta\,\psi_{\alpha\mu} &=& \nabla_\mu \epsilon_\alpha - \frac{1}{4}
T^-_{\rho\sigma} \gamma^{\rho} \gamma^{\sigma} \, \gamma _\mu \,
\varepsilon_{\alpha\beta}\epsilon^\beta \, ,  \nonumber  \label{grtrans} \\
\delta\lambda^{A\alpha} & = & i \, \gamma^\mu \partial_\mu z^A
\epsilon^\alpha + {\cal G}^{-A}_{\rho\sigma}\gamma^{\rho}\gamma^{\sigma}
\varepsilon ^{\alpha\beta} \epsilon_\beta
\end{eqnarray}
where 
\begin{eqnarray}
T_{\mu \nu }^{-} &=&M_{I}F_{\mu \nu }^{I}-L^{I}G_{I\mu \nu }=2i({\hbox{Im}} 
{\cal {N}}_{IJ})L^{I}F_{\mu \nu }^{J-},  \nonumber \\
{\cal G}_{\mu \nu }^{-A} &=&-g^{A\bar{B}}{\bar{f}_{\bar{B}}^{I}}({\hbox{Im}}%
{\cal {N}})_{IJ}F_{\mu \nu }^{J-},  \nonumber \\
G_{I\,\mu \nu }&=&{\hbox{Re}}{\cal {N}}_{IJ}F_{\mu \nu }^{J}-{\hbox{Im}}%
{\cal {N}}_{IJ}{^{\star }}F_{\mu \nu }^{J},  \nonumber \\
\nabla_\mu \epsilon_\alpha&=&(\partial_\mu-{\frac{1}{4}}w^{ab}_\mu\gamma_a
\gamma_b + {\frac{i}{2}}Q_\mu)\epsilon_\alpha
\end{eqnarray}
where $\epsilon_\beta$ is the chiral supersymmetry parameter, $w^{ab}_\mu$
is the spin connection and $Q_\mu$ is the K\"{a}hler connection.

We now review the construction of the stationary solutions to the theory of
ungauged $N=2$ supergravity \cite{black24}. It is well known from the work of Tod \cite{to}
that such metrics can be brought to the following form 
\begin{equation}
ds^2 = - e^{2U} (dt + \omega_m dx^m)^2 + e^{-2 U} dx^m dx^m.
\end{equation}
For this metric, the components of the spin connection are given as follows 
\begin{eqnarray}
w_{t \,0i} &=& - e^{2U} \partial_i U, \qquad w_{t \, ij} = \frac{1}{2}
e^{4U} ( \partial_i w_j - \partial_j w_i),  \nonumber \\
w_{m\, 0i} &=& \frac{1}{2} e^{2U} (\partial_m w_i - \partial_i w_m) - e^{2U}
w_m \partial_i U,  \nonumber \\
w_{m\,ij} &=& \partial_i U \delta_{mj} - \partial_j U \delta_{mi} + \frac{1}{%
2}e^{4U} w_m (\partial_i w_j -\partial_j w_i).
\end{eqnarray}


The equations of motion and the Bianchi identities for the gauge fields can
be solved by 
\[
F_{ij}^{I}={\frac{1}{2}}e^{2U}\varepsilon _{ijm}\partial _{m}{\tilde H}%
^{I}\quad ,\quad G_{I\,ij}={\frac{1}{2}}e^{2U}\varepsilon _{ijm}\partial
_{m}H_{I}\;\;\;, 
\]
where $(\tilde{H}^{I}(x),H_{I}(x))$ are harmonic functions.

The BPS solution can be expressed in the following form \cite{black24} 
\begin{eqnarray}
e^{-2U} &\equiv &Z{\bar{Z}}\equiv i({\bar{Y}}^{I}{\cal {F}}_{I}(Y)-Y^{I} 
\bar{{\cal {F}}}_{I}({\bar{Y}}))  \nonumber \\
\frac{1}{2} e^{2U}\varepsilon_{mnp} \partial_n w_p&\equiv & {\cal Q}_m=
e^{2U} \hbox{Re} (\bar{{\cal {F}}}_{I}(\bar{Y})\partial _{\mu }Y^{I}-{\bar{Y}%
}^{I}\partial _{\mu } {\cal {F}}_{I}(Y),  \nonumber \\
i(Y^{I}-{\bar{Y}}^{I}) &=&i(\bar ZL^{I}-Z{\bar{L}}^{I}) ={\tilde{H}}^{I} 
\nonumber \\
i({\cal {F}}_{I}(Y)-\bar{{\cal {F}}} _{I}(\bar Y)&=&i(\bar ZM_{I}-Z{\bar{M}}%
_{I}) =H_I
\end{eqnarray}
where $Q_{\mu }={\cal Q}_\mu-{\frac{i}{2}}\partial_\mu\log({\frac{\bar Z}{Z}}%
).$

For this particular choice of the metric, we obtain 
\begin{eqnarray}
T_{ij}^{-} &=&{\frac{1}{2}}{\frac{e^{2U}}{\bar{Z}}}\varepsilon
_{ijm}(F_{I}(Y)\partial _{m}{\tilde H}^{I}-Y^{I}\partial _{m}H_{I}), 
\nonumber \\
T_{0k}^{-} &=&\frac{i}{2}\frac{e^{2U}}{\bar{Z}}(F_{I}(Y)\partial _{k}{\tilde
H}^{I}-Y^{I}\partial _{k}H_{I}).
\end{eqnarray}

For the above Ansatz, one can demonstrate that the time-component for the
gravitino supersymmetry transformation as well as those of the gaugino
vanish for the following choice of the spinor supersymmetry parameter 
\begin{equation}
\sqrt{\frac{\bar{Z}}{Z}}\;\epsilon _{\alpha }=i\,\gamma _{0}\epsilon
_{\alpha \beta }\epsilon ^{\beta }.  \label{kilspi}
\end{equation}
Also the equation for the supersymmetry spinor is given by 
\begin{equation}
\left( \partial _{m}+{i}{\cal Q}_{m}+{\frac{1}{2}}\partial _{m}\log {\bar{Z}}%
\right) \epsilon _{\alpha }=0.
\end{equation}
The integrability condition enforces the condition that the field strength
of ${\cal Q}_{m}$ has to vanish. For static non-singular solutions, i.e., $%
w_m=0$, one imposes the vanishing of ${\cal Q}_m.$

\section{Electric BPS solutions In 4D N=2 Anti-De Sitter Supergravity}

In this section we derive the solution for electric BPS states in the theory
of Abelian gauged $N=2$ supergravity coupled to vector supermultiplets \cite
{bigone}. The theory of gauged $N=2$ supergravity without vector multiplets
was first discussed in \cite{das} and BPS solutions of this theory were
discussed in \cite{romans, perry}. More recently, supersymmetric topological
black holes were obtained in the theory of $N=2$ anti-de Sitter supergravity
in \cite{klemm}. Here we will concentrate on electrically charged
spherically symmetric BPS solutions of the theory of gauged $N=2$
supergravity coupled to vector supermultiplets. The Abelian gauging is
achieved by introducing a linear combination of the abelian vector fields $%
A_{\mu }^{I}$ already present in the theory, $A_{\mu }={\kappa }_{I}A_{\mu
}^{I}$ with a coupling constant $g$, where ${\kappa }_{I}$ are constants.
The coupling of the fermi-fields to this vector field breaks supersymmetry
which in order to preserve one has to introduce gauge-invariant $g$%
-dependent terms. In a bosonic background, these additional terms result in
a scalar potential \cite{bigone} 
\begin{equation}
V=g^{2}\left( g^{A\bar{B}}{\kappa }_{I}{\kappa }_{J}f_{A}^{I}\bar{f_{\bar{B}%
}^{J}}-3{\kappa }_{I}{\kappa }_{J}\bar{L}^{I}L^{J}\right) .
\end{equation}

Moreover, the supersymmetry transformations for the gauginos and the
gravitino in a bosonic background become (in terms of complex spinors), 
\begin{eqnarray}
\delta\,\psi_{\mu} &=&\Big({\cal {D}}_\mu +\frac{i}{4} T^-_{\rho\sigma}
\gamma^{\rho} \gamma^{\sigma} \gamma _\mu -ig{\kappa}_IA_\mu^I+{\frac{i}{2}}{%
g}{\kappa}_IL^I \gamma_\mu\Big)\epsilon ,  \label{ggrtrans} \\
\delta\lambda^{A} & = &\Big( i \, \gamma^\mu \partial_\mu z^A
+iG^{-A}_{\rho\sigma}\gamma^{\rho}\gamma^{\sigma} -gg^{A\bar B}{\kappa}%
_If_{\bar B}^I\Big)\epsilon.
\end{eqnarray}

In what follows we are mainly concerned with bosonic backgrounds which break half of
supersyymetry. Consider the following general form for the metric 
\begin{equation}
ds^2=-e^{2A}dt^2+e^{2B}dr^2+e^{2C}r^2(d\theta^2+\sin^2\theta d\phi^2).
\end{equation}
The vielbeins of this metric are 
\begin{eqnarray}
e_t^0&=& e^A, \quad e_r^1=e^B, \quad e_\theta^2=re^C, \quad
e_\phi^3=re^C\sin\theta,  \nonumber \\
e_0^t &=&e^{-A}, \quad e_1^r=e^{-B}, \quad e_2^\theta={\frac{1}{re^C}}\ ,
\quad e_3^\phi={\frac{1}{re^C\sin\theta}}\ .
\end{eqnarray}
and the spin connections 
\begin{eqnarray}
\omega_t^{01}&=&A^{\prime}e^{A-B},  \nonumber \\
\omega_\theta^{12}&=&-(1+rC^{\prime})e^{C-B},  \nonumber \\
\omega_\phi^{13}&=&-(1+rC^{\prime})e^{C-B}\sin\theta,  \nonumber \\
\omega_\phi^{23}&=&-\cos\theta.
\end{eqnarray}

Motivated by the form of the BPS solution in $D=5,$ $N=2$ gauged
supergravity \cite{bcs}, we choose the following Ansatz for the electrically
charged BPS solutions in four-dimensional AdS supergravity 
\begin{eqnarray}
e^{2A}&=&e^{-2B}=e^{2U}f^2, \qquad e^{2C}=e^{-2U},  \nonumber \\
e^{-2U} &\equiv &Z{\bar{Z}}\equiv i({\bar{Y}}^{I}{\cal {F}}_{I}(Y)-Y^{I} 
\bar{{\cal {F}}}_{I}({\bar{Y}})) =Y^IH_I,  \nonumber \\
i(Y^{I}-{\bar{Y}}^{I}) &=&0,  \nonumber \\
i({\cal {F}}_{I}(Y)-\bar{{\cal {F}}} _{I}(\bar Y)&=&H_I,  \nonumber \\
A_t^I&=&e^{2U}Y^I.
\end{eqnarray}
where $U$ and $f$ are functions depending only on the radial distance and
are to be determined.

In terms of the coordinates $(Y^I, F_I(Y))$, the vanishing of the time
component for the gravitino supersymmetry transformation gives the following
equation 
\begin{equation}
\Big(\partial_t-{\frac{1}{2}}(e^{2U}f\partial_r
f+f^2e^U\partial_r(e^U))\gamma_0\gamma_1 +{\frac{f}{4\bar Z}}e^{3U}\partial_r
e^{-2U} \gamma_1 -ige^{2U}{\kappa}_IY^I+{\frac{i}{2}}gf{\kappa}%
_IL^Ie^U\gamma_0\Big)\epsilon=0  \label{tc}
\end{equation}
In what follows we set $Z=\bar Z$. Assuming that the solution breaks
supersymmetry in such a way that the spinor $\epsilon $ satisfies 
\begin{equation}
\epsilon =(a\gamma _{0}+b\gamma _{1})\epsilon  \label{cond}
\end{equation}
where $a$ and $b$ are functions satisfying $a^2-b^2=1$ and to be determined.
The condition (\ref{cond}) breaks half of supersymmetry. From Eq. (\ref{tc})
one obtains the following equations 
\begin{eqnarray}
\Big(\partial_t+{\frac{a}{2b}}e^{2U}{f\partial_r f}-ige^{2U}{\kappa}_IY^I\Big)%
\epsilon&=&0,  \nonumber \\
{a}f\partial_r e^U+{\frac{1}{2}}e^{3U}\partial_r(e^{-2U})&=&0,  \nonumber \\
-{\frac{1}{b}}e^{U}\partial_r f+{ig}{\kappa}_IL^I+fb\partial_r e^U&=&0.
\label{pet}
\end{eqnarray}
The gaugino supersymmetry transformation is given by 
\begin{equation}
\delta\lambda^{A} = \Big( i \, \gamma^\mu \partial_\mu z^A +i{\cal G}%
^{-A}_{\rho\sigma}\gamma^{\rho}\gamma^{\sigma} -gg^{A\bar B}{\kappa}%
_If_{\bar B}^I\Big)\epsilon  \label{fin}
\end{equation}
where ${\cal G}_{\rho \nu }^{-\,A}=-g^{A\bar{B}}\,\bar{f}_{\bar{B}%
}^{I}\,\left( {\hbox{Im}}{\cal {N}}_{IJ}\right) F_{\ \rho \nu }^{-\,J}$, $%
g^{A\bar{B}}$ is the inverse K\"{a}hler metric and $\bar{f}_{\bar{B}%
}^{I}=(\partial _{\bar{B}}+{\frac{1}{2}}\partial _{\bar{B}}K)\bar{L}^{I}$.
To show the vanishing of the gaugino supersymmetry variations for the choice
of $\epsilon$ given in (\ref{cond}), it is more convenient to multiply Eq. (%
\ref{fin}) with $f^I_A$. This gives using the second relation in (9)
following from special geometry 
\begin{equation}
f_A^I\delta \lambda^{\alpha A} =\Big( i \gamma^{\mu} \partial_{\mu}
z^A(\partial_A + {\frac{1}{2}} \, \partial_A K) L^I +{\frac{i }{2}} (F^{- \,
I}_{\mu\nu} - i \bar{L}^I T^-_{\mu\nu}) \gamma^{\mu} \gamma^{\nu}+{\frac{g}{2%
}}{\hbox{Im}}{\cal {N}}^{IJ}{\kappa}_J+g{\kappa}_JL^J{\bar L}^I\Big)\epsilon
\end{equation}

We now try to simplify the above rather complicated form by evaluating each
term separately. One has 
\begin{eqnarray}
\partial _{\mu }z^{A}\left[ (\partial _{A}+{\frac{1}{2}}\,\partial
_{A}K)L^{I}\right]  &=&e^{\frac{K}{2}}\partial _{\mu }X^{I}-\partial _{\mu
}z^{A}(\partial _{A}e^{-K})e^{\frac{3K}{2}}X^{I} \\
&=&e^{\frac{K}{2}}\left[ \partial _{\mu }X^{I}-iX^{I}(\bar{X}^{J}\partial
_{\mu }F_{J}-\partial _{\mu }X^{J}\bar{F}_{J})e^{K}\right] 
\end{eqnarray}
which for our Ansatz becomes 
\begin{equation}
\partial _{\mu }z^{A}(\partial _{A}+{\frac{1}{2}}\,\partial
_{A}K)L^{I}=-e^{3U}Y^{I}({Y}^{J}\partial_r H_{J})+e^{U}\partial_r Y^{I}.
\end{equation}
The second term in the gaugino transformation gives 
\begin{equation}
{\frac{i}{2}}\Big (F_{\mu \nu }^{-\,I}-i\bar{L}^{I}T_{\mu \nu }^{-})\gamma
^{\mu }\gamma ^{\nu }\epsilon =-2i\Big(F_{01}^{-\,I}-i{\bar{L}}^{I}T_{01}^{-}%
\Big)\gamma _{0}\gamma _{1}\epsilon .
\end{equation}
Using the supersymmetry breaking condition (\ref{cond}) 
\begin{equation}
\gamma _{0}\gamma _{1}\epsilon =-\Big({\frac{\gamma _{1}}{a}}+{\frac{b}{a}}%
\Big)\epsilon 
\end{equation}
we get 
\begin{equation}
{\frac{i}{2}}\Big (F_{\mu \nu }^{-\,I}-i\bar{L}^{I}T_{\mu \nu }^{-})\gamma
^{\mu }\gamma ^{\nu }\epsilon ={\frac{2}{a}}\Big(iF_{01}^{-\,I}\gamma
_{1}+ibF_{01}^{-\,I}+{\bar{L}}^{I}T_{01}^{-}\gamma _{1}+b{\bar{L}}%
^{I}T_{01}^{-})\epsilon .
\end{equation}

Collecting terms and imposing the vanishing of the gaugino supersymmetry
transformation gives the following equations 
\begin{eqnarray}
\Big(-{ie^{4U}f}Y^{I}({Y}^{J}\partial_r H_{J})+{ie^{2U}f}\partial_r Y^I\Big) +{%
\frac{2}{a}}{\bar L}^IT^{-}_{01}+{\frac{2i}{a}}F^{-\, I}_{01}&=&0, 
\nonumber \\
{\frac{2 b}{a}}{\bar L}^IT^-_{01}+{\frac{2ib}{a}}F^{-\, I}_{01}+{\frac{g}{2}}{%
\hbox{Im}}{\cal {N}}^{IJ}{\kappa}_J+g\kappa_JL^J{\bar L}^I&=&0.
\label{boring}
\end{eqnarray}
It can be shown that all the conditions imposed by unbroken supersymmetry which are given by (%
\ref{boring}) and (\ref{pet}) are satisfied if we set 
\begin{eqnarray}
a &=&{\frac{1}{f}},  \nonumber \\
b &=&-{i\frac{gre^{-2U}}{f}},  \nonumber \\
f^{2} &=&1+g^{2}r^{2}e^{-4U}
\end{eqnarray}
and take the harmonic functions to be $H_I={\kappa}_I +{%
\frac{q_I}{r}}$, where $q_I$ are electric charges.

Here we summarize our solution 
\begin{eqnarray}
& &ds^2=-\Big(e^{2U}+g^2r^2e^{-2U}\Big)dt^2+{\frac{1}{\Big(%
e^{2U}+g^2r^2e^{-2U}\Big)}} dr^2+e^{2U}r^2(d\theta^2+\sin^2\theta d\phi^2), 
\nonumber \\
& &e^{-2U}=Y^IH_I,  \nonumber \\
& &i({\cal {F}}_{I}(Y)-\bar{{\cal {F}}}_{I}(\bar Y)=H_I, \qquad Y^{I}={\bar{Y%
}}^{I}  \nonumber \\
& &A_t^I=e^{2U}Y^I.
\end{eqnarray}

For the above solution, the vanishing of the time-component of the gravitino
supersymmetry variation simply gives 
\begin{equation}
\Big(\partial _{t}-i{\frac{g}{2}}\Big)=0.
\end{equation}
From the vanishing of the space-component of the gravitino supersymmetry
transformation we obtain the following equations
\begin{eqnarray}
\Big(\partial _{r}+{\frac{1}{2fr}}(1-2r\partial _{r}U)\gamma _{0}-{\frac{1}{2r%
}}(1-r\partial _{r}U)\Big)\epsilon  &=&0,  \nonumber \\
\Big(\partial _{\theta }+{\frac{1}{2}}\gamma _{0}\gamma _{1}\gamma _{2}\Big)%
\epsilon  &=&0,  \nonumber \\
\Big(\partial _{\phi }+{\frac{1}{2}}\cos \theta \gamma _{2}\gamma _{3}+{%
\frac{1}{2}}\gamma _{0}\gamma _{1}\gamma _{3}\sin \theta \Big)\epsilon  &=&0.
\nonumber \\
&&
\end{eqnarray}

The radial differential equation can be solved using the techniques
described in the Appendix of \cite{romans}. If one has for the spinor $%
\Psi(r)$ the following differential equation 
\begin{equation}
\partial_r\Psi(r)=\Big(A(r)+B(r)\Gamma_1\Big)\Psi(r),
\end{equation}
where $\Gamma_1$ is an operator satisfying $\Gamma_1^2=1.$ Also suppose that $%
\Psi(r)$ is subject to the constraint 
\begin{equation}
\Psi(r)=-\Big(X(r)\Gamma_1+Y(r)\Gamma_2\Big)\Psi(r), \qquad \Gamma_2^2=1,\quad \{\Gamma_1,\Gamma_2\}=0,
\end{equation}
then integrability implies that 
\begin{equation}
{\frac{dX}{dr}}+2BY^2=0,  \label{ran}
\end{equation}
and the solution for $\Psi(r)$ is given by 
\begin{equation}
\Psi(r)={\frac{1}{2}} \Big(V(r)+W(r)\Gamma_2\Big)(1-\Gamma_1)\Psi_0,
\end{equation}
where 
\begin{eqnarray}
V(r)&=&\sqrt{{\frac{1+X}{Y}}}e^T,  \nonumber \\
W(r)&=&-\sqrt{{\frac{1-X}{Y}}}e^T  \nonumber \\
T& =&\int^r A(r^{\prime})dr^{\prime}
\end{eqnarray}
and $\Psi_0$ is a constant arbitrary spinor. For the case at hand, we have
the following identifications\footnote{%
notice that the integrability condition (\ref{ran}) is satisfied.} 
\begin{eqnarray}
\Gamma_1&=&\gamma_0,\qquad \Gamma_2=i\gamma_1,  \nonumber \\
A(r)&=& {\frac{1}{2r}}(1-r\partial_rU),  \nonumber \\
B(r)&=&-{\frac{1}{2rf}}(1-2r\partial_rU),  \nonumber \\
X(r)&=&-{\frac{1}{f}}  \nonumber \\
Y(r)&=&{\frac{gre^{-2U}}{f}}
\end{eqnarray}

By combining with the solution for the differential equations for time and
the angular variables, we obtain the following solution 
\begin{equation}
\epsilon (r)={\frac{1}{2\sqrt{gr}}}e^{\frac{igt}{2}}e^{-{\frac{1}{2}}\gamma
_{0}\gamma _{1}\gamma _{2}\theta }e^{-{\frac{1}{2}}\gamma _{2}\gamma
_{3}\phi }e^{U+T}\Big(\sqrt{f-1}-i\gamma _{1}\sqrt{f+1}\big)(1-\gamma
_{0})\epsilon _{0}
\end{equation}
where $\epsilon _{0}$ is an arbitrary constant spinor and ${T}=\int^{r}{%
\frac{1}{2r^{\prime }}}(1-r^{\prime }\partial _{r^{\prime }}U(r^{\prime
}))dr^{\prime }$.

In summary, we have obtained spherically symmetric electric BPS solutions
which break half of supersymmetry in the theory of $N=2$ anti-De Sitter
supergravity with vector supemultiplets. The solution is expressed in terms
of the holomorphic sections and therefore independent of the existence of a
holomorphic prepotential. Note that a subclass of solutions of $N=2$
supergravity (for a particular choice of prepotential) are actually also
solutions of supergravity theories with more, i.e. $N=4$ or $N=8$
supersymmetries. The D=4 static BPS-saturated electric black holes of gauged
supergravity have naked singularities. The generalisation of the results of
this paper to the non-extreme static black hole solutions is straightforward
and can be done along the same lines of \cite{cams}. \vfill\eject
\newpage 

\noindent {\Large {\bf Appendix: conventions and notation}} We took the
space--time metric to have signature $(-+++)$. Curved indices were denoted
by $\mu = t, m$. Flat indices were denoted by $a=0, i$ where $i=1,2,3$.
Antisymmetrized indices were defined as follows: $[ab] = {\frac{1 }{2}} (ab
-ba)$. We defined (anti) selfdual components as follows: 
\begin{equation}
F^{\pm}_{ab} = {\frac{1 }{2}}(F_{ab} \pm i\, {^{\star}F_{ab}})
\end{equation}
with 
\begin{equation}
{^{\star}F^{ab}} = {\frac{1 }{2}} \varepsilon^{abcd}F_{cd}
\end{equation}
and $\varepsilon^{0123} = 1 = - \varepsilon_{0123}$ ($^{\star\star}F= -F$).
For the $\gamma$ matrices we used the relation 
\begin{equation}
\gamma^a \gamma^b = - \eta^{ab} + {\frac{i }{2}} \gamma_5 \epsilon^{abcd}
\gamma_c \gamma_d
\end{equation}
with $\gamma_5 = -i \gamma_0 \gamma_1 \gamma_2 \gamma_3$ ($\gamma_5^2 =1$).
Using these definitions we find for any antiselfdual tensor the identity 
\begin{equation}
T^{-\, ab} \gamma_a \gamma_b = 2(1 - \gamma_5) T^{-\, 0m} \gamma_0 \gamma_m
\end{equation}
and 
\begin{equation}
T^-_{mn} = - i\, \epsilon_{mnp} T^-_{0p} \ .
\end{equation}
\vfill\eject


\begin{thebibliography}{99}
\bibitem{first}  S. Ferrara, R. Kallosh and A. Strominger, {\it Phys. Rev.} 
{\bf D52} (1995) 5412; S. Ferrara and R. Kallosh, {\it Phys. Rev.} {\bf D54}
(1996) 1514; S. Ferrara, R. Kallosh {\it Phys. Rev.} {\bf D54} (1996) 1525.

\bibitem{black24}  K. Behrndt and W. A. Sabra, {\it Phys. Lett.} {\bf B401}
(1997) 258; W. A. Sabra, {\it Mod. Phys. Lett}. {\bf A12}(1997) 2585; {\it %
Nucl. Phys.} {\bf B 510} (1998) 247; K. Behrndt, D. L\"{u}st and W. A.
Sabra, {\it Nucl. Phys.} {\bf B510} (1998) 264;  K. Behrndt, G. L. Cardoso, B. de Wit, D. L\"ust, T. Mohaupt 
and W. A. Sabra, {\it Phys. Lett. }  {\bf B429} (1998) 289.
\bibitem{black25}  W. A. Sabra,\ {\it Mod. Phys. Lett}. {\bf A 13} (1998)
239 ; A. Chamseddine and W. A. Sabra, {\it Phys. Lett}. {\bf B426} (1998)
36.

\bibitem{Strominger}  A. Strominger, {\it Nucl. Phys}. {\bf B451} (1995) 96;
B. R. Greene, D. R. Morrison and A. Strominger, {\it Nucl. Phys}. {\bf B451}
(1995) 109; E. Witten, {\it Nucl. Phys.} {\bf B471} (1996) 195.

\bibitem{topological}  A. Chou, R. Kallosh, J. Rahmfeld, S.-J. Rey, M.
Shmakova, W. K. Wong; {\it Nucl. Phys.}. {\bf B 508} (1997) 147. K. Behrndt,
D. L\"{u}st and W. A. Sabra, {\it Phys. Lett}. {\bf B 418} (1998) 303; I.
Gaida, S. Mahapatra, T. Mohaupt and W. A. Sabra, {\it Class. Quant. Grav.} 
{\bf 16} (1999) 419.

\bibitem{ibnsina}  J.~Maldacena, {\it Adv. Theor. Math. Phys.}{\bf 2} (1998)
231; S.~S. Gubser, I.~R. Klebanov, and A.~M. Polyakov, {\em Phys. Lett.} 
{\bf B428} (1998) 105; E.~Witten, {\it Adv. Theor. Math. Phys.} {\bf 2}
(1998) 253.

\bibitem{top} R. B. Mann, {\it Class. Quantum Grav.}{\bf 14%
} (1997) L109; \ D. Klemm and L. Vanzo, {\it Phys. Rev. } {D58} (1998)
104025.

\bibitem{romans}  L.~J. Romans, {\em Nucl. Phys.} {\bf B383} (1992) 395.

\bibitem{bcs}  K. Behrndt, A. H. Chamseddine and W. A. Sabra, {\it Phys.
Lett.} {\bf B442} (1998) 97

\bibitem{jim}  M. J. Duff and J. T. Liu, hep-th/9901149.

\bibitem{klemm}  M. M. Caldarelli and D. Klemm, hep-th/9808097.

\bibitem{v}  B. Craps, F. Roose, W. Troost and A. Van Proeyen, {\it Nucl.
Phys.}. {\bf B503} (1997) 565.

\bibitem{to}  K. P. Tod, {\it Phys. Lett.} {\bf B121} (1983) 241; {\it %
Class. Quantum. Grav.} {\bf 12} (1995) 1801.

\bibitem{bigone}  L. Andrianopoli, M. Bertolini, A. Ceresole, R. D' Auri, S.
Ferrara, P. Fre\'{ } and T. Magri, {\it J. Geom. Phys,} {\bf 23} (1997) 111.

\bibitem{das}  A. Das and D. Z. Freedman, {\it Nucl. Phys.} {\bf B120}
(1977) 221.

\bibitem{perry}  V. A. Kostelecky and M. J. Perry, {\it Phys. Lett.} {\bf %
B371} (1996) 191.

\bibitem{cams}  K. Behrndt, M. Cvetic and W. A. Sabra, hep-th/9810227
\end{thebibliography}
\end{document}